\documentclass{aa}
\usepackage{txfonts}
\usepackage{natbib}
\usepackage{graphicx}
\bibpunct{(}{)}{;}{a}{}{,}

\begin{document}
\title{On the emergence of toroidal flux tubes: general dynamics and comparisons with the cylinder model}

\author{D. MacTaggart \and A.W. Hood}

\institute{School of Mathematics and Statistics, University of St Andrews, North Haugh, St Andrews, Fife, KY16 9SS, Scotland}

\abstract {} {In this paper we study the dynamics of toroidal flux tubes emerging from the solar interior, through the photosphere and into the corona.  Many previous theoretical studies of flux emergence use a twisted cylindrical tube in the solar interior as the initial condition.  Important insights can be gained from this model, however, it does have shortcomings.  The axis of the tube never fully emerges as dense plasma becomes trapped in magnetic dips and restrains its ascent.  Also, since the entire tube is buoyant, the main photospheric footpoints (sunspots) continually drift apart.  These problems make it difficult to produce a convincing sunspot pair.  We aim to address these problems by considering a different initial condition, namely a toroidal flux tube.}{We perform numerical experiments and solve the 3D MHD equations.  The dynamics are investigated through a range of initial field strengths and twists.}{The experiments demonstrate that the emergence of toroidal flux tubes is highly dynamic and exhibits a rich variety of behaviour.  In answer to the aims, however, if the initial field strength is strong enough, the axis of the tube can fully emerge.  Also, the sunspot pair does not continually drift apart.  Instead, its maximum separation is the diameter of the original toroidal tube.} {}

\keywords{Sun: magnetic fields - Magnetohydrodynamics (MHD) - Methods: numerical}

\titlerunning{On the emergence of toroidal flux tubes}
\authorrunning{D. MacTaggart et al.}
\maketitle

\section{Introduction}
  It is generally accepted that solar active regions, and therefore sunspots, are the result of magnetic flux tubes that have risen from the base of the convection zone and emerged at the photosphere.  The emerging flux tubes are $\Omega$-shaped, at least near photosphere, and produce bipolar sunspot pairs.  This has been the classical picture of sunspot formation for some time.  \citet{cowling46} envisaged flux tubes running as girdles around the Sun and suggested that loops were carried upwards by convection to emerge as sunspot pairs.  Soon after, the idea of \emph{magnetic buoyancy} was put forward as the mechanism by which flux tubes journey from the base of the convection zone to the photosphere \citep{parker55,jensen55}.  Parker demonstrated that an isolated magnetic flux tube in a stratified plasma under gravity must be buoyant, provided that it is close to thermal equilibrium with its surroundings.  Although the exact mechanism by which flux tubes rise in the convection zone is not known, most theorists tend to favour magnetic buoyancy.  Simulations using spherical geometry have now been performed which show how a twisted flux tube can survive rising buoyantly though the convection zone \citep{jouve07,fan08}. 

Due to the wide variety of phenomena that exist in sunspots, different modelling techniques have been developed to describe them.  These range from equilibrium models \citep[e.g.][]{petrie99} to studies of magnetoconvection \citep[e.g.][]{heinemann07}.  A useful collection of these methods, and general sunspot physics, is presented in \citet{tw92,tw08}.  In recent years it has become possible to simulate the rise of flux tubes in the upper convection zone and their emergence into the atmosphere.  The first 3D simulation of the emergence of a twisted flux tube in a stratified atmosphere was performed by \citet{fan01}.  Several similar simulations have been performed since and are described in a recent review by \citet{archontis08rev}.  What all of these types of flux emergence simulations have in common is that they use the same initial condition, namely a cylindrical twisted flux tube in the solar interior.  To encourage the formation of an $\Omega$-shaped loop, a density deficit is introduced into the tube that is proportional to $e^{-y^2/\lambda^2}$, where $y$ is the horizontal coordinate along the tube length and $\lambda$ is a parameter describing the length of the buoyant section of the tube.  When the $\Omega$-loop emerges it creates a sunspot pair at the photosphere.  Much has been gained from this model, however, it contains some drawbacks.  The cylindrical tube continues to emerge  and the sunspot pair continually drifts apart.  This is not seen in observations, where active regions spread to a finite size and then decay \citep{liu06}.  Another drawback is that, in simulations with this initial condition, the original axis of the tube remains trapped at the base of the photosphere and never emerges to be almost vertical when it leaves and enters the sunspots.  In this paper we consider a different geometry for initial the magnetic field - a \emph{toroidal flux tube}.  This model was originally considered by \citet{hood09}.  We shall investigate the behaviour of emerging toroidal tubes through a range of initial field strengths and twists.  Our approach is to consider an idealized stratified solar interior and atmosphere.  We are interested in the interaction of the plasma and the magnetic field, so we solve the 3D resistive and compressive magnetohydrodynamic (MHD) equations.  The aim is, using the toroidal model, to answer the following questions:

\begin{enumerate} 
\item Can the axis of the original flux tube emerge into the solar atmosphere and, if so, how?
\item Can the two main polarities of the active region (sunspots) drift to a fixed distance and then stop?
\end{enumerate}
Throughout the paper, comparisons will be made with the well-studied cylindrical model.  The outline of the paper is as follows: $\S 2$ will describe the model setup and the initial conditions used in the flux emergence simulations.  In $\S 3$ we will examine the dynamics of toroidal emergence over a range of parameters.  First, a general description of the emergence process is explained.  This is followed by a more detailed examination of particular cases.  $\S 4$ will summarize the conclusions.

\section{Model setup}
In this section we outline the numerical setup and the initial atmosphere and magnetic field configurations.

\subsection{Main equations}
For our numerical experiments we use a 3D version of the Lagrangian remap scheme detailed in \citet{arber01}.  This code has been used for other flux emergence studies \citep{archontis08,archontis09}.  For each time step, the equations are solved in Lagrangian form and are then remapped back onto an Eulerian grid.  All of the physics is contained in the Lagrangian step.  The variables are made dimensionless against photospheric values.  These values are: pressure, $p_{\mathrm{ph}} = 1.4\times 10^{4}\,\, \mbox{Pa}$; density, $\rho_{\mathrm{ph}} = 3\times 10^{-4}\,\,\mbox{kg}\,\,\mbox{m}^{-3}$; temperature, $T_{\mathrm{ph}} = 5.6\times 10^3\,\, \mbox{K}$ and scale height $H_{\mathrm{ph}} = 170 \,\,\mbox{km}$.  The other units used in the simulations are: time, $t_{\mathrm{ph}} = 25\,\,\mbox{s}$; speed, $u_{\mathrm{ph}} = (p_{\mathrm{ph}}/\rho_{\mathrm{ph}})^{1/2} = 6.8\,\,\mbox{km}\,\,{s}^{-1}$ and magnetic field $ B_{\mathrm{ph}} = 1.3\times 10^3\,\,\mbox{G}$.  
The evolution of the system is governed by the following time-dependent and resistive (non-dimensionalized) MHD equations:

\begin{equation}
 \frac{\partial \rho}{\partial t} + \nabla\cdot (\rho\mathbf{u}) = 0, 
\end{equation}
\begin{equation}
\rho\left(\frac{\partial\mathbf{u}}{\partial t} + (\mathbf{u}\cdot\nabla)\mathbf{u}\right) = -\nabla p + (\nabla\times\mathbf{B})\times\mathbf{B} + \nabla\cdot\mathbf{T} + \rho\mathbf{g},
\end{equation}
\begin{equation} 
\frac{\partial\mathbf{B}}{\partial t} = \nabla\times(\mathbf{u}\times\mathbf{B}) + \eta\nabla^2\mathbf{B}, \\
\end{equation}
\begin{equation}
\rho\left(\frac{\partial\varepsilon}{\partial t} + (\mathbf{u}\cdot\nabla)\mathbf{\varepsilon}\right) = -p\nabla\cdot\mathbf{u} + \eta j^2 + Q_{\mathrm{visc}},
\end{equation}
\begin{equation}\label{solenoid}
 \nabla\cdot\mathbf{B} = 0,
\end{equation}

with specific energy density

\[
 \varepsilon = \frac{p}{(\gamma-1)\rho}.
\]
The basic variables are the density $\rho$, the pressure $p$, the magnetic field vector $\mathbf{B}$ and the velocity vector $\mathbf{u}$. $j$ is the magnitude of current density and $\mathbf{g}$ is gravity (uniform in the $z$-direction). $\gamma$ is the ratio of specific heats and is taken as 5/3.  $\eta$ is the resistivity which is taken to be uniform. For all the experiments we take $\eta = 0.001$. The viscosity tensor is
\[
 \mathbf{T} = \mu\left(\nabla\mathbf{u} + \nabla\mathbf{u}^{\mathrm{T}} - \frac{2}{3}\mathbf{I}\,\nabla\cdot\mathbf{u}\right),
\]
where $\mu$ is the viscosity and $\mathbf{I}$ is the identity tensor.  The contribution of viscosity to the energy equation is represented by $Q_\mathrm{visc}$.  The difference in the results of the simulations with and without (small) compressive viscosity is found to be negligible.  The code accurately resolves shocks by using a combination of artificial viscosity and Van Leer flux limiters.  In such regions, heating is added to the energy equation.  The equations are solved on a uniform Cartesian grid $(x,y,z)$ of $(128,128,256)$ for the (non-dimensionalized) region $-50\le x \le 50$, $-50\le y \le 50$ and $-25\le x \le 85$.  The boundary conditions are periodic on the side walls of the computational box and the top and bottom boundaries are closed.  A damping layer is included at the top of the box to reduce the reflection of waves. 

\subsection{Initial atmosphere}
The initial stratification of the atmosphere is similar to that used in previous flux emergence studies \citep{fan01,murray06,dmac09}.  The solar interior $(z\leq 0)$ is taken to be marginally stable to convection since in this study we are focussing on the emerging field.  The effects of convection are left for future work.  The photosphere/chromosphere lies in the region $0\leq z \leq 10$, the transition region in $10 \leq z \leq 20$ and the corona in $z \geq 20$.  The (non-dimensionalized) temperature is specified as

\[
T(z) = \left\{\begin{array}{cc}
1-z\frac{\gamma-1}{\gamma} & \quad z\le 0, \\
1 &  \quad 0 < z \le 10, \\
T_{\mathrm{cor}}^{(z-10)/10} &  \quad 10 < z \le 20, \\
T_{\mathrm{cor}} & z > 20,
\end{array} \right.
\] 
where $T_{\mathrm{cor}}=150$ is the coronal temperature.  The other state variables, pressure and density, are found by numerically solving the hydrostatic equation.  This gives a numerically stable equilibrium.  

\subsection{Initial magnetic field}
As mentioned earlier, nearly all previous studies of numerical flux emergence use a twisted cylindrical flux tube in the solar interior as the initial condition.  Here we present the cylinder and the toroidal models.

\subsubsection{Cylinder model}
The magnetic field of a twisted cylindrical flux tube \citep{fan01} is given by

\begin{equation}\label{cylinder}
 \mathbf{B} = B_y(r)\hat{\mathbf{y}} + B_{\theta}(r)\hat{\mathbf{\theta}},
\end{equation}
where
\begin{eqnarray}
\label{by} 
B_y(r) &=& B_0e^{-r^2/r_0^2}, \\
\label{btheta} 
B_{\theta}(r) &=& \alpha r B_y(r).
\end{eqnarray}
$\hat{\mathbf{y}}$ is the direction of the tube axis and $\hat{\mathbf{\theta}}$ is the azimuthal direction in the tube cross-section. $r_0$ is the radius of the tube and $r^2 = x^2 + z^2$.  The flux tube is uniformly twisted with $\alpha$ denoting the angle of field line rotation about the axis over a unit length of the tube.  $B_0$ is the initial field strength at the axis of the tube.

The pressure inside the tube differs from the surrounding field-free region by $p_{\mathrm{def}}$.  Balancing the radial components of the Lorentz force and the plasma pressure gradient gives

\begin{equation}\label{cylinder2}
 \frac{B_\theta}{r}\frac{{\rm d}}{{\rm d}r}(rB_\theta) + \frac{1}{2}\frac{{\rm d}B_y^2}{{\rm d}r} + \frac{{\rm d}p_{\mathrm{def}}}{{\rm d}r} = 0.
\end{equation}
Solving this gives a pressure deficit, relative to the background hydrostatic pressure, of
\begin{equation}
 p_{\mathrm{def}}(r) = B_0^2e^{-2r^2/r_0^2}(\alpha^2r_0^2-2-2\alpha^2r^2)/4.
\end{equation}
The density deficit can then be calculated from $\rho_{\mathrm{def}} = p_{\mathrm{def}}/T(z)$.  As mentioned before, to encourage an $\Omega$-loop, $\rho_{\mathrm{def}}$ is multiplied by the factor $e^{-y^2/\lambda^2}$ to make the middle of the tube more buoyant.  To our knowledge, all previous flux emergence studies using the (buoyant) cylinder model form the $\Omega$-loop using this exponential profile.  Some studies impose a velocity profile on the cylinder model to make it rise and emerge, rather than make the tube buoyant \citep[e.g.][]{magara03}.

\subsubsection{Toroidal model}
A full derivation of this model is given in \citet{hood09} so we shall only outline certain steps here. If, in Cartesian coordinates, the tube axis lies in the $y$-direction then the equations for the magnetic field can be written in polar coordinates $(s,\phi,x)$:

\[
s^2 = y^2 + (z-z_0)^2 \quad \mbox{with} \quad s\cos\phi = y \quad \mbox{and} \quad s\sin\phi = z-z_0,
\]
where $z_0$ is the base of the computational box.  Under the assumption of rotational invariance, we take the magnetic field to be of the form 

\begin{eqnarray*}
\mathbf{B} &=& \nabla A \times \nabla\phi + B_\phi \mathbf{e}_\phi \\
&=& -\frac{1}{s}\frac{\partial A}{\partial x}\mathbf{e}_s + \frac{1}{s}\frac{\partial A}{\partial s}\mathbf{e}_x + B_\phi \mathbf{e}_\phi,
\end{eqnarray*}
where $A$ is the flux function and is constant on magnetic field lines.  This form automatically satisfies the solenoidal constraint, equation (\ref{solenoid}).  After some manipulation, insertion of this field into the magnetostatic balance equation, $(\nabla\times\mathbf{B})\times\mathbf{B} = \nabla p_{\mathrm{def}}$, yields the Grad-Shafranov equation

\begin{equation}\label{gs}
 \frac{\partial^2 A}{\partial s^2} - \frac{1}{s}\frac{\partial A}{\partial s} +\frac{\partial^2 A}{\partial x^2} + b_\phi\frac{{\rm d} b_\phi}{{\rm d}A} + s^2\frac{{\rm d}p_{\mathrm{def}}}{{\rm d}A} = 0,
\end{equation}
where $b_\phi = sB_\phi$.  We now define a local toroidal coordinate system $(r,\theta,\phi)$, where

\[
 r^2 = x^2 + (s-s_0)^2 \quad \mbox{with} \quad s-s_0 = r\cos\theta \quad \mbox{and} \quad x = r\sin\theta,
\]
with major axis $s_0$.  Rewriting equation (\ref{gs}) in these local coordinates and then taking a regular expansion in the aspect ratio $r_0/s_0$ produces the leading order balance equation 

\[
 \frac{B_\theta}{r}\frac{{\rm d}}{{\rm d}r}(rB_\theta) + \frac{1}{2}\frac{{\rm d}B_\phi^2}{{\rm d}r} + \frac{{\rm d}p_{\mathrm{def}}}{{\rm d}r} = 0.
\]
This has the same form as equation (\ref{cylinder2}) from the cylinder model.  We can, therefore, choose the solutions to be the same as those for the straight tube.  These are, as in (\ref{by}) and (\ref{btheta}), 

\[
 B_\phi = B_0e^{-r^2/r_0^2}, \quad B_\theta = \alpha r B_\phi = \alpha B_0 r e^{-r^2/r_0^2}.
\]
Again, $B_0$ is the axial field strength and $\alpha$ is the twist.  The pressure difference is, again, $p_{\mathrm{def}}(r) = B_0^2e^{-2r^2/r_0^2}(\alpha^2r_0^2-2-2\alpha^2r^2)/4.$  The temperature profile is specified and the density deficit is given by
\begin{equation}\label{rho_def}
\rho_{\mathrm{def}} =  B_0^2e^{-2r^2/r_0^2}(\alpha^2r_0^2-2-2\alpha^2r^2)/(4T(z)).
\end{equation}
In the simulations the entire toroidal tube is made buoyant.  We could find more terms in the expansion but since we do not require an exact equilibrium (as the tube will be made buoyant) we only consider the leading order solution.

The resulting magnetic field for a twisted toroidal tube is given (in Cartesian coordinates) by

\begin{eqnarray}
B_x &=& B_\theta (r)\frac{s-s_0}{r}, \\
B_y &=& -B_\phi (r)\frac{z-z_0}{s} - B_\theta(r)\frac{x}{r}\frac{y}{s}, \\
B_z &=& B_\phi(r)\frac{y}{s} - B_\theta (r)\frac{x}{r}\frac{z-z_0}{s}.
\end{eqnarray}
\section{Parameter study}
Now that the basic model is established, we shall investigate the dynamics of toroidal flux emergence by considering a range of parameters.  In this study we look at the effects of changing the initial field strength $B_0$ and the initial twist $\alpha$.  The other parameters will be kept constant in this paper: $s_0=15$, $r_0=2.5$ and $z_0=-25$.
\subsection{Varying $B_0$ with fixed $\alpha$: general dynamics}
In this section we consider the effects of changing the initial field strength, $B_0$, and keep the initial twist fixed at $\alpha = 0.2$.  This value is smaller than those used in previous studies \citep{fan01, murray06, archontis08, hood09} and is believed to be more applicable to the Sun.  We follow the evolution for the cases $B_0=1,3,5,7$ and $9$.
\begin{figure}
 \resizebox{\hsize}{!}{\includegraphics{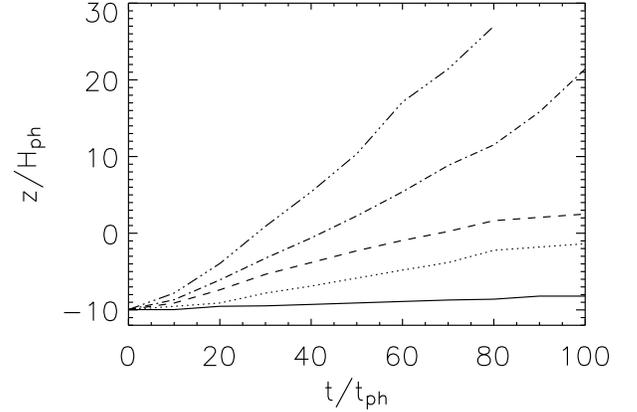}}
 \caption{The height-time profiles for axes with different $B_0$ traced in the $y=0$ plane. Key: $B_0=1$ (solid), $B_0=3$ (dot), $B_0=5$ (dash), $B_0=7$ (dot-dash) and $B_0=9$ (triple dot-dash).}
\label{axis_heights}
\end{figure}
\begin{figure}
 \resizebox{\hsize}{!}{\includegraphics{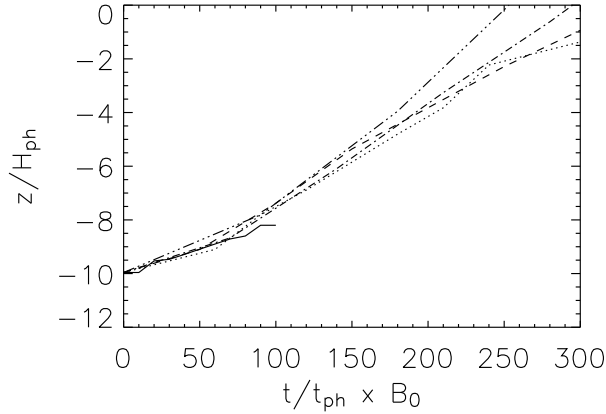}}
 \caption{The height-time profiles in the solar interior rescaled to time $\bar{t} = tB_0$.  Key: $B_0=1$ (solid), $B_0=3$ (dot), $B_0=5$ (dash), $B_0=7$ (dot-dash) and $B_0=9$ (triple dot-dash).}
\label{axis_heights_similar}
\end{figure}

From equation (\ref{rho_def}) it can be seen that the buoyancy force on the tube is proportional to $B_0^2$.  It is then expected that tubes with a stronger $B_0$ will rise faster and further than those with a smaller initial value.  This is confirmed in the simulations and the height-time profiles of the tube axes are displayed in Figure \ref{axis_heights}.  The axes are tracked by examining the change in $B_x$ in the $y=0$ plane.  The field line structure is also examined to confirm that the the axes do indeed pass through the plane at the change in $B_x$.  This method is applicable since the tubes are weakly twisted.  By rescaling the time as $\bar{t} = tB_0$, the heights reached by the axes are similar in the solar interior (see Figure \ref{axis_heights_similar}).  This is equivalent to measuring time on the Alfv\'{e}n timescale rather than a sound timescale.  Thus the heights of the tube axes are not only a function of time but also of initial field strength, i.e. $H(\bar{t}) = H(tB_0)$ where $H$ is the height function of a tube axis.  This agrees with the behaviour of cylindrical tubes as found by \citet{murray06}.   The two low field strength cases, $B_0=1,3$, do not reach the photosphere in the time the simulation is run.  These tubes may emerge if their evolution is tracked for a much longer time span.  The axes of the other three cases, $B_0=5,7,9$, all emerge above the base of the photosphere.  These cases differ from the cylindrical model, where the tube axis remains trapped near the base of the photosphere $(z=0)$.  There is also a distinction between the cases themselves.  The magnetic field evolution of the strong field cases, $B_0=7,9$, differs from that of the moderate field case, $B_0=5$.  Their axes rise to the corona whereas the axis for the $B_0=5$ case stops in the middle of the photosphere.  Before we consider why this occurs we will describe the general behaviour of the emergence of toroidal flux tubes.

\begin{figure}
 \resizebox{\hsize}{!}{\includegraphics{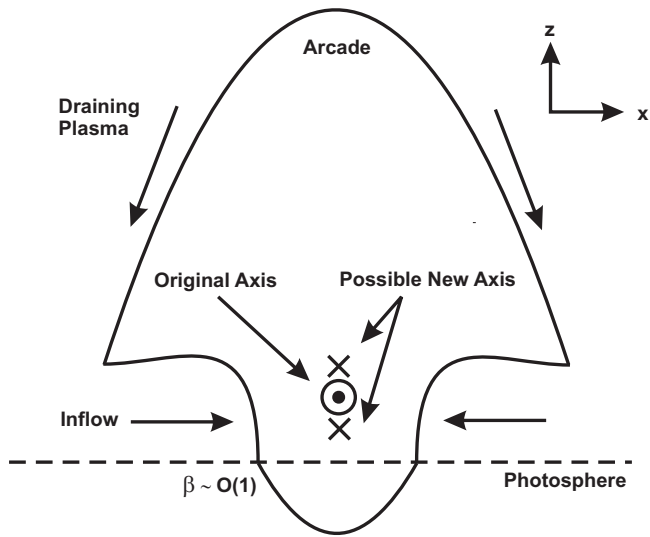}}
 \caption{This diagram illustrates some of the main dynamical features of the emergence of a toroidal flux tube.  The outline represents the outermost field line of the emerging arcade. See the text for more details.}
\label{cartoon}
\end{figure}

Figure \ref{cartoon} displays a diagram showing some of the main features of the emergence process based on the dynamics found in the simulations.  When the flux tube reaches the base of the photosphere $(z=0)$, the plasma beta $(\beta = p/(|\mathbf{B}|^2/2))$ decreases until it is $O(1)$.  There the magnetic field becomes subject to a magnetic buoyancy instability \citep{murray06} and expands rapidly into the solar atmosphere which has an exponential decrease in pressure with height.  As the magnetic arcade expands, the magnitude of $B_y$ decreases with height due to flux conservation.  This gradient in $B_y$ results in a Lorentz force that drives horizontal shear flows along the neutral line between the bipolar sources \citep{manchester01,manchester04}.  This can be understood by considering the $y$-component of the tension from the Lorentz force:
\[
\{(\mathbf{B}\cdot\nabla)\mathbf{B}\}_y = \mathbf{B}\cdot\nabla B_y. 
\]
The gradient of $B_y$ is negative moving in the direction of $\mathbf{B}$ on one side of the arcade and is positive moving in the direction of $\mathbf{B}$ on the other.  Horizontal shearing occurs from the base of the photosphere to the top of the arcade.

As the magnetic field expands into the atmosphere, plasma drains from the top of the arcade and follows field lines down to the photosphere.  Due to the rapid expansion of the arcade, a pressure and density deficit forms at the centre of the emerging region just above the photosphere.  As the pressure at the centre of the arcade is smaller than that further out at the photosphere, the plasma, drained from the top of the arcade, flows into this region (see Figure \ref{cartoon}).  The plasma either collects there or drains down the legs of the toroidal tube.

The combination of horizontal shearing and inflow can bring together inclined field lines and initiate magnetic reconnection.  This process can result in the formation of new flux ropes in the solar atmosphere.  However, the new axis will be either above or below the original axis, as shown in Figure \ref{cartoon}.

\subsection{$B_0=5,7$ comparison}

Now that a basic description of the dynamics of flux emergence has been presented we shall examine, more closely, particular cases that exhibit different classes of behaviour.  As shown before (see Figure \ref{axis_heights}) the original tube axis of the $B_0=5$ case stops rising in the middle of the the photosphere, whereas the axis of the $B_0=7$ case rises to the corona.  We shall now compare these cases.

\begin{figure}
 \resizebox{\hsize}{!}{\includegraphics{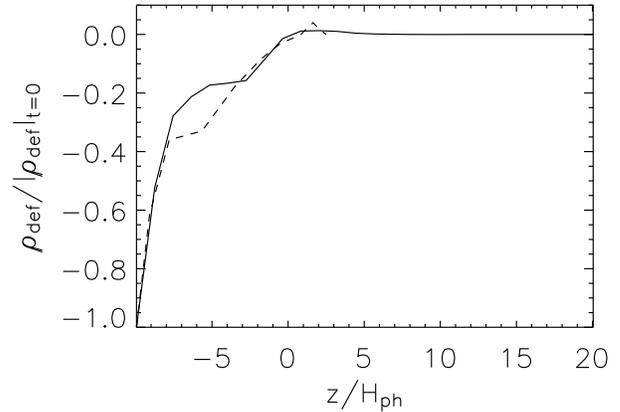}}
 \caption{The density deficit at the tube axis (in the $y=0$ plane) as a fraction of the initial unsigned density deficit against height.  Key: $B_0=5$ (dash), $B_0=7$ (solid).}
\label{rho_heights}
\end{figure}

\begin{figure}
 \resizebox{\hsize}{!}{\includegraphics{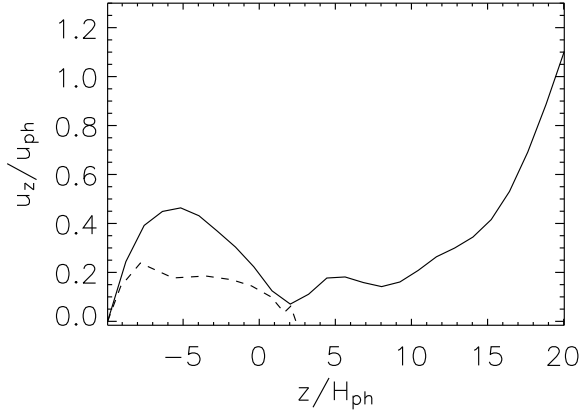}}
 \caption{The vertical velocity, $u_z$, (in the $y=0$ plane) as a function of height.  $u_{\mathrm{ph}}$ is the photospheric velocity scale and $H_{\mathrm{ph}}$ is the photospheric scale height. Key: $B_0=5$ (dash), $B_0=7$ (solid).}
\label{vel_heights}
\end{figure}

Figure \ref{rho_heights} shows how the density deficit at the axis, as a fraction of the initial unsigned density deficit, varies with height.  The deficit is calculated by taking density values from inside and outside the tube, at the same height, in the $y=0$ plane.  This gives a measure of the buoyancy of the tubes.   The evolution of the two cases is similar.  Both curves rise until the tubes become neutrally buoyant at $z \approx 2.5$.  The $B_0=7$ case becomes neutrally buoyant at an earlier time than the $B_0=5$ case because the stronger field case rises faster.  The point of neutral buoyancy corresponds to $t \approx 110$ for the $B_0=5$ case and $t \approx 52$ for the $B_0=7$ case.  One important point not note is that both cases become neutrally buoyant in the photosphere.  This is not found for the cylindrical model where the tubes become over dense \emph{before} reaching the photosphere \citep{murray06}.

Figure \ref{vel_heights} displays how the vertical velocity, $u_z$, at the axis varies with height.  The $B_0=7$ case achieves more than double the rise velocity of the $B_0=5$ case in the solar interior.  Both cases initially follow a similar profile. i.e. $u_z$ increases until a maximum is reached and then decreases.  For the $B_0=5$ case, the velocity decreases to approximately zero at a height $z = 2.4$.  This corresponds to a time of $t \approx 110$.  At this height, the $B_0=7$ case has a rise velocity of $u_z \approx 0.1$, at time $t \approx 52$ demonstrating that the initial choice of $B_0$ is crucial in determining the the evolution of the axis properties.  However, the story is more complicated as the axis heights in the photosphere and above are also influenced by draining flows.  

\begin{figure}
 \resizebox{\hsize}{!}{\includegraphics{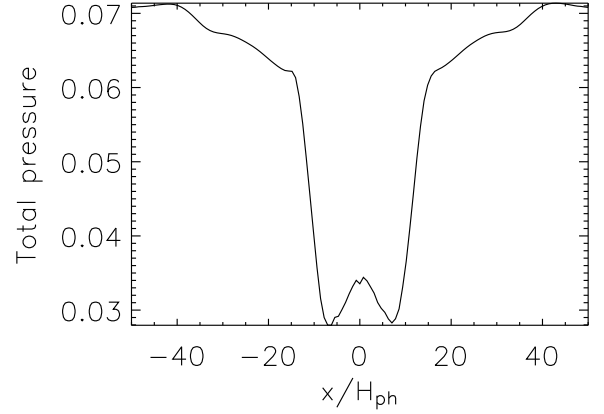}}
 \caption{The `square well' profile for the deficit in the total pressure. This deficit exists for a finite range of heights (see text) and plasma draining from above flows into it.  This figure is for the $B_0=7$ case at $t=100$ and $(y,z)=(0,3)$.}
\label{pressure}
\end{figure}

\begin{figure}
 \resizebox{\hsize}{!}{\includegraphics{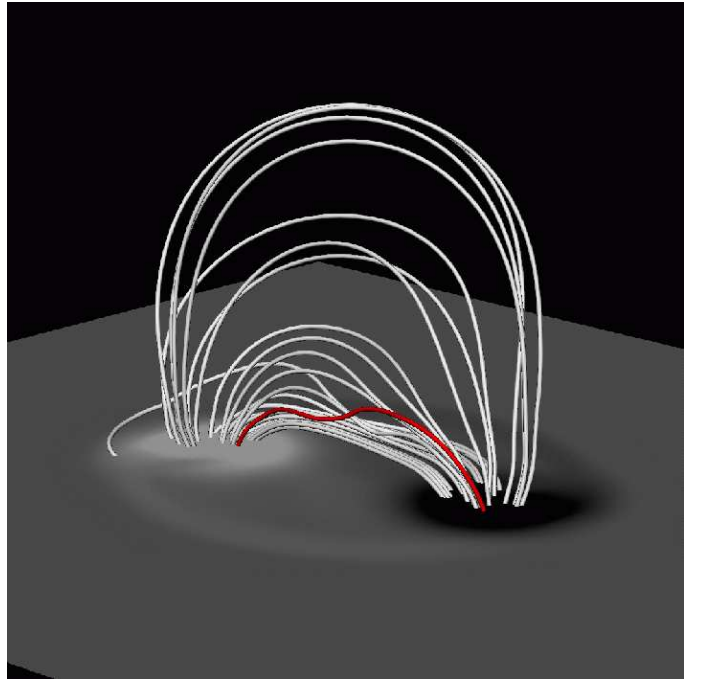}}
 \caption{The $B_0=5$ case at $t=100$.  The original tube axis is represented by a red field line.  Some surrounding field lines are traced in grey.  A magnetogram is placed at the bottom of the photosphere (z=0) and shows $B_z$.}
\label{b0_5_axis}
\end{figure}

\begin{figure}
 \resizebox{\hsize}{!}{\includegraphics{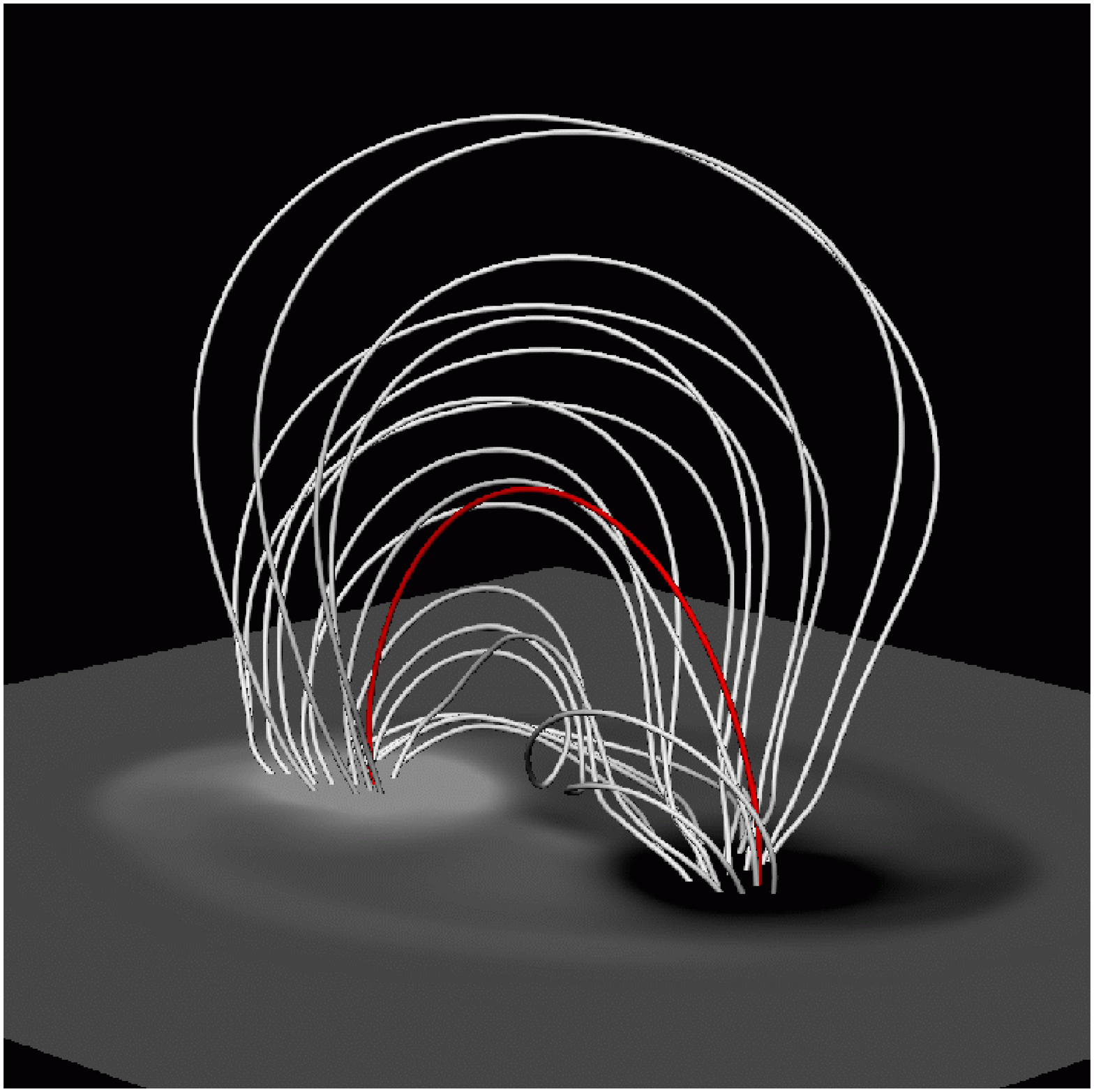}}
 \caption{The $B_0=7$ case at $t=100$.  The original tube axis is represented by a red field line.  Some surrounding field lines are traced in grey.  A magnetogram is placed at the bottom of the photosphere (z=0) and shows $B_z$.}
\label{b0_7_axis}
\end{figure}

As previously described, plasma drains down the emerged magnetic arcade and then flows into a region of reduced total pressure $(p + |\mathbf{B}|^2/2)$.  An example of this region from the $B_0=7$ case is displayed in Figure \ref{pressure}.This \lq square well\rq$\,$ profile exists between the heights of $z\approx 1.64$ and $z\approx 6.8$ for the $B_0=5$ case and $z\approx 1.64$ and $z\approx 5.1$ for the $B_0=7$ case.  For $B_0=5$, the original tube axis rises slowly (compared with the $B_0=7$ case) and just reaches the bottom of the pressure deficit region when plasma flows into it.  It is this plasma that flows on \emph{top} of the original axis and prevents its further ascent.  Figures \ref{b0_5_axis} and \ref{b0_7_axis} illustrate the positions of the original tube axis, for both cases, in relation to the field structure of the magnetic arcade at $t=100$.

\begin{figure}
 \resizebox{\hsize}{!}{\includegraphics{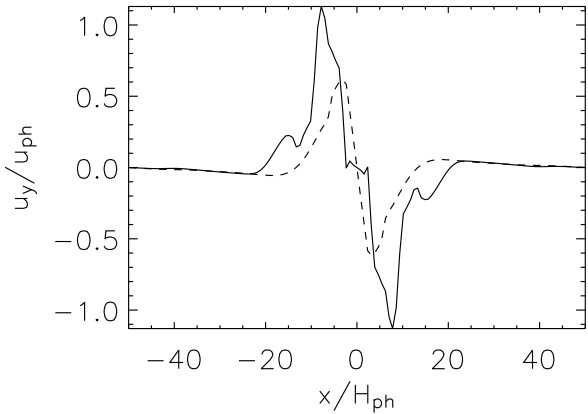}}
 \caption{Shearing profiles at $t=100$ and $(y,z)=(0,3)$ for $B_0=5$ (dash) and $B_0=7$ (solid).}
\label{shear}
\end{figure}

\begin{figure}
 \resizebox{\hsize}{!}{\includegraphics{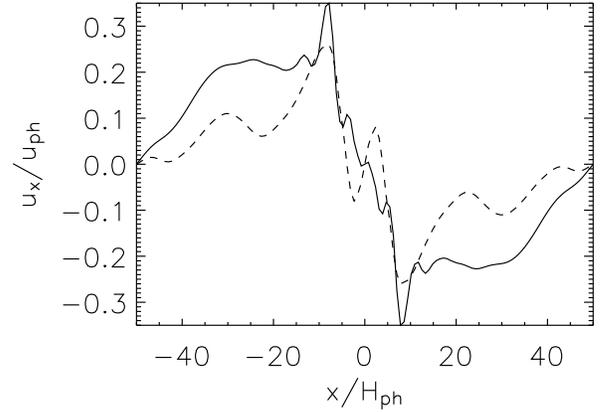}}
 \caption{Inflow profiles at $t=100$ and $(y,z)=(0,3)$ for $B_0=5$ (dash) and $B_0=7$ (solid). There are clear inflow profiles for both cases.  However, when the flows meet in the centre of the pressure deficit zone, they produce more complex behaviour.}
\label{inflow}
\end{figure}

\begin{figure}
 \resizebox{\hsize}{!}{\includegraphics{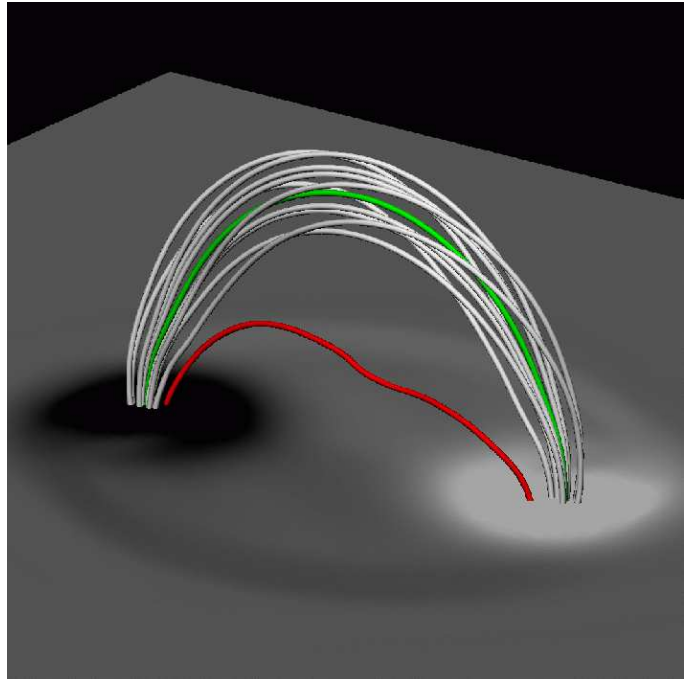}}
 \caption{The $B_0=5$ case at $t=100$.  The red field line represents the original tube axis.  The green field line represents the axis of a new flux rope. The surrounding field line structure at the new axis is demonstrated by some field lines traced in grey.  The original axis is pinned down in the photosphere whereas the new rope is at the base of the corona.  The magnetogram is at $z=0$.}
\label{axis1}
\end{figure}

\begin{figure}
 \resizebox{\hsize}{!}{\includegraphics{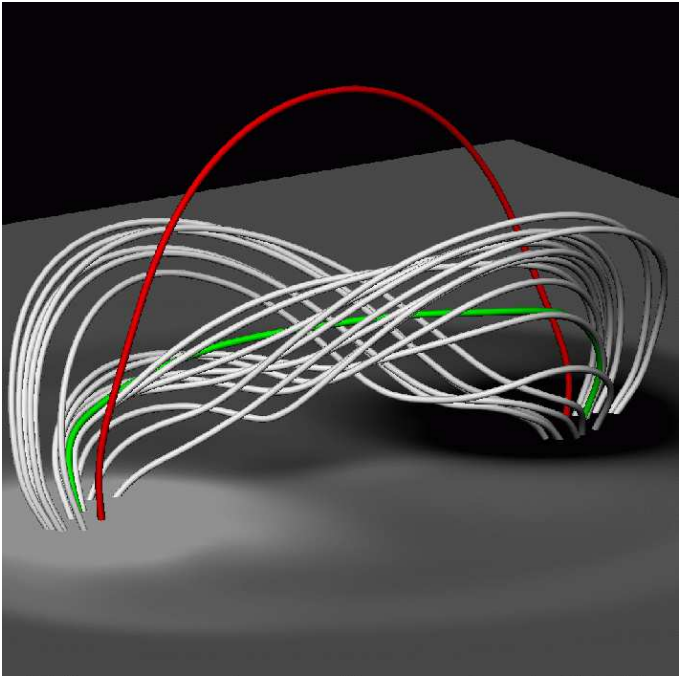}}
 \caption{The $B_0=7$ case at $t=100$.  The red field line represents the original tube axis.  The green field line represents the axis of a new flux rope. The surrounding field line structure at the new axis is demonstrated by some field lines traced in grey.  Reconnection occurs where the grey field lines cross.  An upward jet from this pushes the original axis higher.  The magnetogram is at $z=0$.}
\label{axis2}
\end{figure}

As previously described, there is horizontal shearing ($y$-direction) in the arcade as it expands.  This motion combined with inflowing plasma can lead to magnetic reconnection.  Figures \ref{shear} and \ref{inflow} shows examples of shearing and inflows for both cases, respectively.  The combination of these flows does indeed lead to reconnection, in both simulations, and results in the formation of new flux ropes.  In the $B_0=5$ case, a flux rope forms above the original axis and is able to rise to the corona.  Figure \ref{axis1} shows the field line structure of the new rope for the $B_0=5$ case in relation to the axis of the original tube.  Similar behaviour has been observed in simulations using the cylinder model \citep{manchester04,archontis08}.  In the $B_0=7$ case,  the original axis emerges to the corona and a new flux rope forms below it.  To our knowledge this has not been found in previous theoretical flux emergence studies that use the cylinder model and do not include convection.  The flux rope forms directly below the original axis and the reconnection that creates it produces an upflow.  This upflow gives an extra kick to the rising of the original axis and explains the steep increase in the $u_z$ curve for $B_0=7$ in Figure \ref{vel_heights}.  The field line structure of the new flux rope in the $B_0=7$ case is shown in Figure \ref{axis2}.

\begin{figure}
 \resizebox{\hsize}{!}{\includegraphics{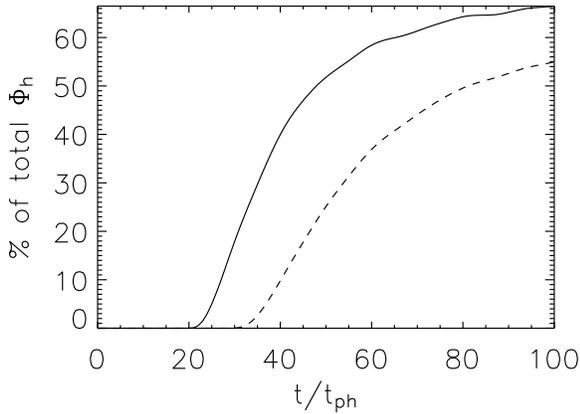}}
 \caption{The evolution of $\Phi_h(y=0,t)/\Phi_h(y=0,t=0)\times 100\%$ for $B_0=5$ (dash) and $B_0=7$ (solid) in the atmosphere.}
\label{above}
\end{figure}

\begin{figure}
 \resizebox{\hsize}{!}{\includegraphics{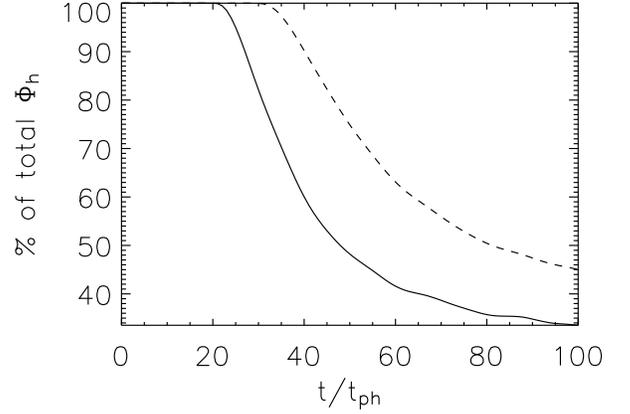}}
 \caption{The evolution of $\Phi_h(y=0,t)/\Phi_h(y=0,t=0)\times 100\%$ for $B_0=5$ (dash) and $B_0=7$ (solid) in the interior.}
\label{below}
\end{figure}

In the emergence process not all the flux is transported into the atmosphere, some remains in the solar interior and at the base of the photosphere.  To quantify how much flux emerges and how much does not we consider the horizontal flux, through the central $y=0$ plane,

\[
 \Phi_h(y=0) = \int\int B_y\, {\rm d}x {\rm d}z.
\]
This integral is calculated for the regions above and below the base of the photosphere $(z=0)$.  These values are shown in Figures \ref{above} and \ref{below} as percentages of the initial $\Phi_h$, through time, for $B_0=5$ and $B_0=7$.  As described above, the $B_0=7$ case rises faster and emerges before the $B_0=5$ case.  Also, the stronger field case transports more flux into the atmosphere, as expected.  The horizontal flux remains constant, for both cases, when the flux tube rises in the interior and has not yet reached the photosphere.  When the tube reaches the phototsphere and becomes subject to the buoyancy instability, horizontal flux is transported into the atmosphere and the same amount is depleted in the interior.

\subsection{Plasma drainage}\label{pd}
We have answered the first part of question one, which was asked in the Introduction, by showing that the original axis of a toroidal tube, with sufficiently strong $B_0$, can emerge to the corona.  However, the problem of why this happens in the toroidal model and not in the standard cylindrical model remains to be confronted.  \cite{murray06} carry out a parameter study for the cylinder model.  For a tube of $B_0=9$ they find the maximum height of the original tube axis reaches $z\approx 2$.  The main difference between the two models lies in the geometry.  The legs of the toroidal model rise steeply as the whole arch of the tube rises to emerge.  The cylinder model, however, with its exponential buoyancy profile, kinks in the centre of the tube.  The buoyant section increases as the tube rises higher. 

\begin{figure}
 \resizebox{\hsize}{!}{\includegraphics{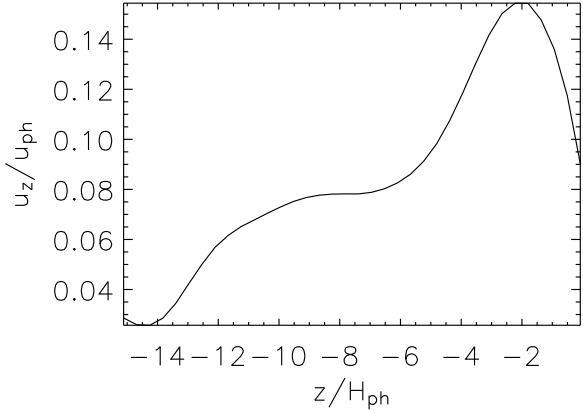}}
 \caption{The vertical velocity profile through a cut in one of the legs of the toroidal tube at $t=40$ and $(x,y)=(0,-13)$.  The vertical velocity in the cut is positive since the tube is buoyantly rising.}
\label{rise}
\end{figure}

\begin{figure}
 \resizebox{\hsize}{!}{\includegraphics{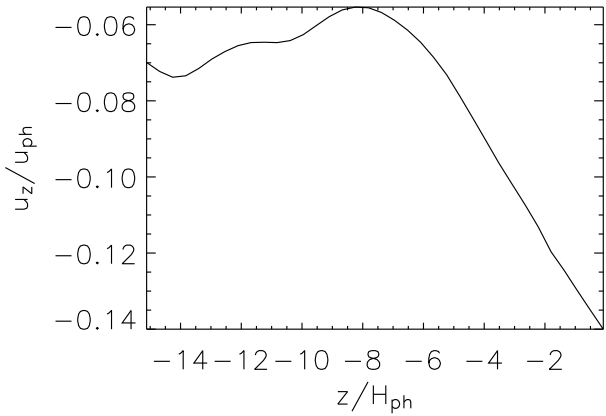}}
 \caption{The vertical velocity profile through a cut in one of the legs of the toroidal tube at $t=50$ and $(x,y)=(0,-13)$.  The sign of the velocity in the cut has now changed since plasma drains down the leg, creating a downflow.}
\label{drain}
\end{figure}

As the cylinder model emerges, plasma draining from the emerged arcade flows down to the photosphere and collects in multiple dips where the axis is trapped \citep{archontis09}.  In the toroidal model, flows exist in the legs of the tubes that correspond to draining downflows.  Figures \ref{rise} and \ref{drain} show a cut of $u_z$ through one of the legs for the $B_0=7$ case at $t=40,50$, respectively.  At $t=40$ the tube is buoyantly rising and the vertical velocity in the cut is positive.  By $t=50$, however, plasma begins to drain down the arcade and also down the legs of the tube.  There is a change in sign in the velocity of the cut and this corresponds to a draining downflow.  The geometry of the toroidal model allows the plasma to drain down the legs and not collect in dips.  It is this property that allows the original axis of the tube to emerge.

To determine whether or not the axis can emerge in the cylinder model, we perform a simple test.  Instead of using the the standard buoyancy profile of $\exp(-y^2/\lambda^2)$, we consider 
\[
n\exp(-y^2/\lambda^2) - (n-1), 
\]
where $n$ is a positive integer.  This is a generalisation of the standard profile $(n=1)$ and will make the central part of the tube buoyant and the ends of the tube over dense.  The size of the buoyant region is controlled by $n$ and $\lambda$.  The reason for choosing this profile is to produce a toroidal-like geometry from the cylinder model.  When the experiment begins, the centre will rise and the ends sink, giving the required shape.  For our experiment we choose $n=6$, $\lambda=20$, $B_0=7$ and $\alpha=0.4$.  This strong twist is used to help prevent the breaking up of the tube in the solar interior \citep{emonet98}.  The base of the computational box is lowered to $z_0=-50$ to allow the tube to develop a toroidal profile.  Figure \ref{cylinder} depicts the shape of the tube at $t=86$ by showing an isosurface of $|\mathbf{B}|=0.5$ and a field line indicating the tube axis.

\begin{figure}
 \resizebox{\hsize}{!}{\includegraphics{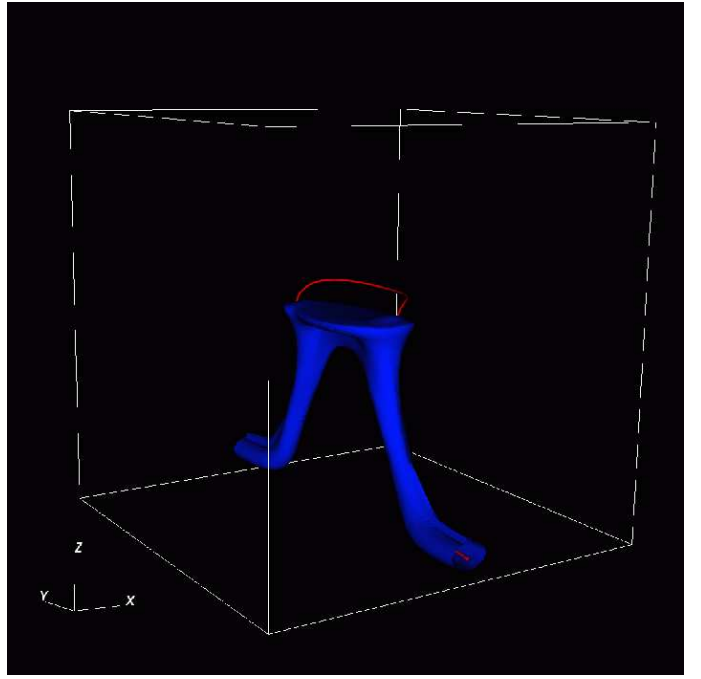}}
 \caption{The geometry of the cylinder model with the buoyancy profile $n\exp(-y^2/\lambda^2) - (n-1)$ exhibits a toroidal-like shape.  This snapshot is at $t=86$ and shows an isosurface of $|\mathbf{B}|=0.5$ in blue and a field line in red indicating the axis.}
\label{cylinder}
\end{figure}

As mentioned at the beginning of this section, \cite{murray06} found that the axis of a tube with $B_0=9$ and $\alpha=0.4$ rises to a maximum height of $z\approx 2$.  In our experiment the, now arched, cylinder axis has risen far beyond this.  By $t=86$ the height of the axis is $z\approx 10$, the base of the transition region.  This confirms that it is the geometry of the toroidal model that enables the efficient draining of plasma and so allows the axis to emerge.  

This may help to explain why flux emergence studies that include convection \citep{cheung07,tortosa09} find that the axis of the cylindrical tube emerges.  If convective flows can  change the geometry of the tube from cylindrical to toroidal, then the axis can emerge as in our experiment.

\subsection{Varying $\alpha$ with fixed $B_0$}

In this section we investigate the effect of varying the initial twist $\alpha$.  We will consider the evolution of tubes with $\alpha$ = 0.2, 0.3, 0.4.  We will take $B_0=7$ for all these cases since, as described in the previous section, this value results in the tube axis emerging with the field around the axis being vertical at the centre of the sunspots.  All tubes have an initial minor radius $r_0=2.5$ and major radius $s_0=15$.

\subsubsection{Rise and emergence}

\begin{figure}
 \resizebox{\hsize}{!}{\includegraphics{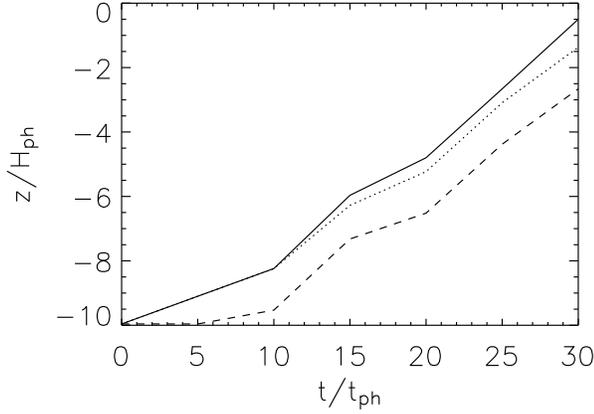}}
 \caption{The axis height-time profiles for the cases $\alpha=0.2$ (solid), $\alpha=0.3$ (dot) and $\alpha=0.4$ (dash) in the solar interior.}
\label{alpha_interior}
\end{figure}

The twist of the magnetic field of a flux tube results in a tension force acting on the tube.  At the top of the initial toroidal tube the density deficit at the axis is given by

\[
 \rho_{\mathrm{def}} = \frac{B_0^2(\alpha^2r_0^2-2)}{4T(z)}. 
\]
The smaller the value of $\alpha$ the larger the deficit.  Hence the axes of tubes with smaller values of $\alpha$ are more buoyant than tubes with larger values.  Although the whole tube is made buoyant, the top of the tube is more buoyant since $T(z)$ monotonically decreases with height in the solar interior.  Figure \ref{alpha_interior} shows the height-time profiles for the tube axes in the solar interior.  As expected, the lower the twist, the faster the rise.

\begin{figure}
 \resizebox{\hsize}{!}{\includegraphics{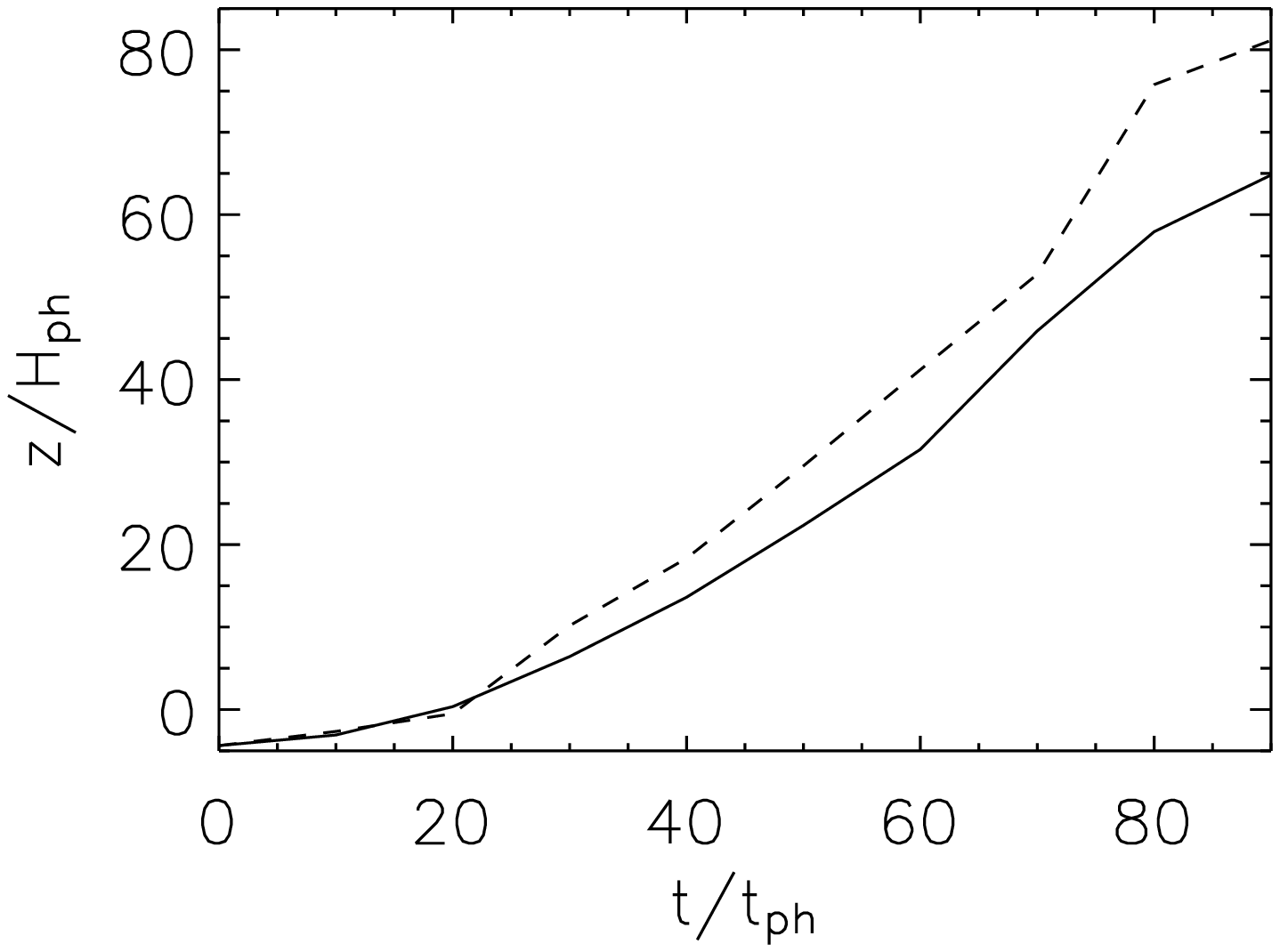}}
 \caption{The amount of twist affects the amount of flux that emerges into the atmosphere.  Key: $\alpha=0.2$ (solid), $\alpha=0.4$ (dash).}
\label{field_top}
\end{figure}

\begin{figure}
 \resizebox{\hsize}{!}{\includegraphics{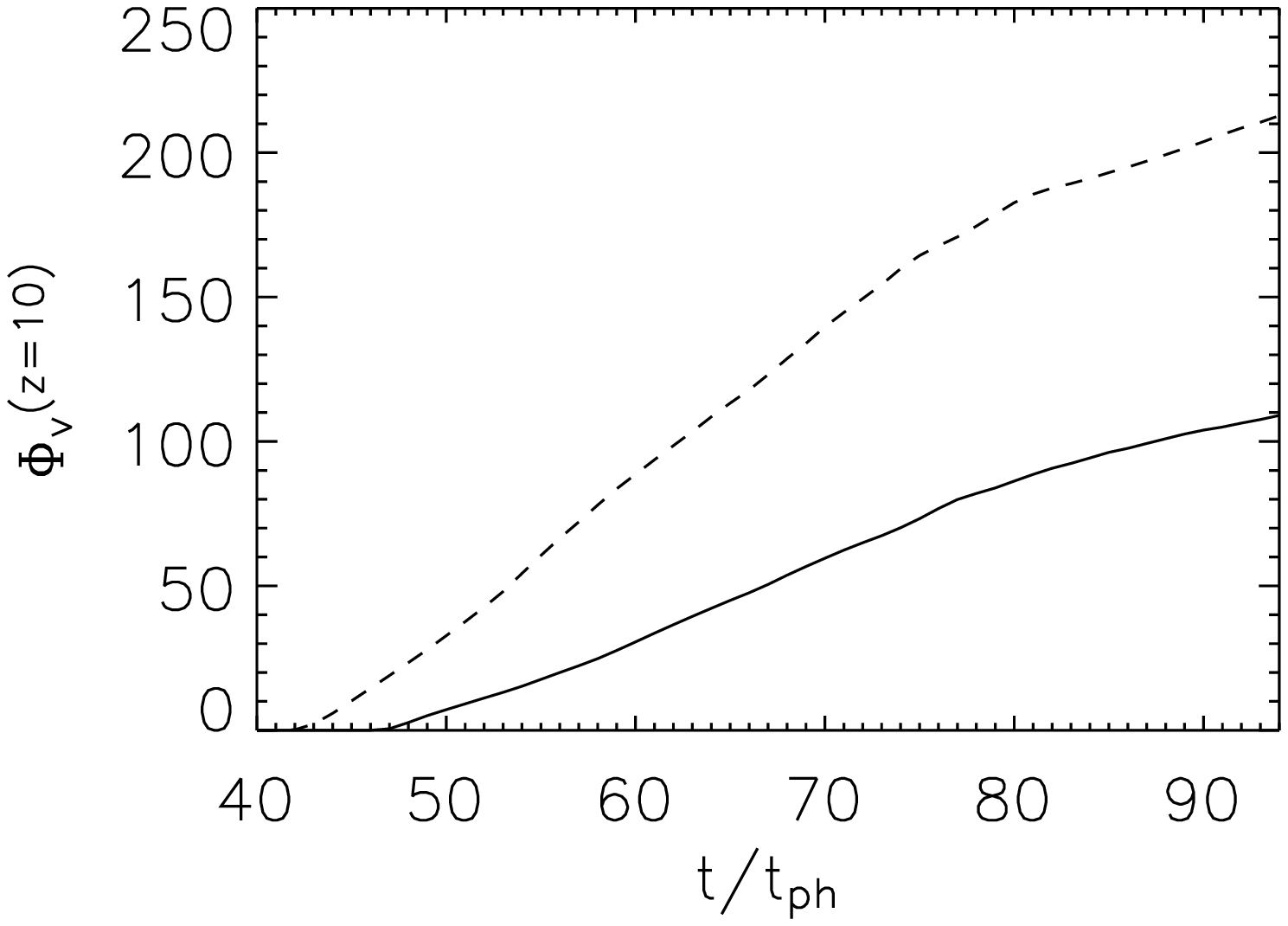}}
 \caption{The amount of twist affects the amount of flux that emerges into the atmosphere.  Key: $\alpha=0.2$ (solid), $\alpha=0.4$ (dash).}
\label{vertical_flux}
\end{figure}

Once the tube reaches the photosphere, it slows down and then emerges by means of a buoyancy instability \citep{murray06}.  To investigate how the initial twist influences the emergence, we look at how the unsigned vertical flux evolves with time.  The total unsigned vertical flux in a plane $(z=z_0)$ is defined by

\[
 \Phi_v(z_0) = \int\int|B_z|\,{\rm d}x{\rm d}y.
\]
Here we will consider the $\alpha=0.2$ and $\alpha=0.4$ cases. Figure \ref{field_top} shows the height-time profile for the top of the magnetic field. i.e. the top of the tube and its expansion into the atmosphere. Figure \ref{vertical_flux}  displays the evolution of $\Phi_v$ for the different twist cases in the plane $z=10$ (base of the transition region).

The $\alpha=0.2$ case rises to the photosphere before the $\alpha=0.4$ case.  At the photosphere, however, the smaller twist case takes longer to become subject to the buoyancy instability than the larger one.  This means that both cases emerge at approximately the same time.  \citet{murray06} found similar behaviour for the cylinder model.  They studied the buoyancy instability  by considering a local analysis at the photosphere.  Once emerged, the field of the $\alpha=0.4$ case rises faster and further than the $\alpha=0.2$ case.  This is shown in Figure \ref{field_top}.  Figure \ref{vertical_flux} shows the evolution of $\Phi_v$ for $\alpha=0.4$ begins before that of $\alpha=0.2$.  It also rises at a faster rate and by $t=94$ there is double the amount of unsigned vertical flux compared with the $\alpha=0.2$ case.  When at the photosphere, the field for the $\alpha=0.2$ case was weaker than that of the $\alpha=0.4$ case.  Therefore, the field emerging into the atmosphere is also weaker, resulting in a weaker evolution of the total unsigned flux.  

\subsubsection{Sunspot drift}

\begin{figure}
 \resizebox{\hsize}{!}{\includegraphics{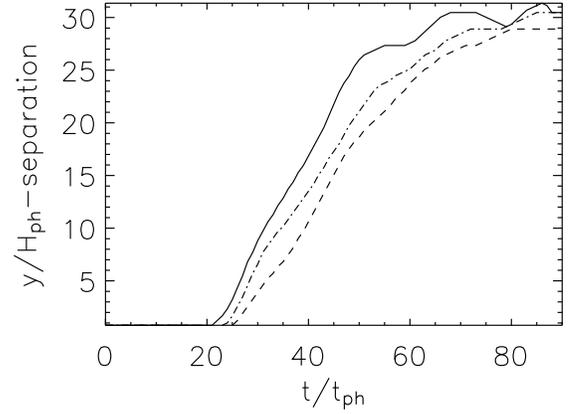}}
 \caption{The $y$-separation in time of the maximum and minimum $B_z$ at the base of the photosphere (z=0).  The higher the twist, the slower the separation. Key: $\alpha=0.2$ (solid), $\alpha=0.3$ (dot-dash), $\alpha=0.4$ (dash).}
\label{footpoints}
\end{figure}

A simulation drawback of the cylinder model, which has become the standard in numerical flux emergence, is that the sunspot pair it produces continually drifts apart until it reaches the edge of the computational box.  Although a density deficit is introduced into the tube to form an $\Omega$-loop, the entire tube is in fact buoyant since the $\exp(-y^2/\lambda^2)$ profile makes the ends of the tube weakly buoyant.  This is what causes the sunspots to drift apart.  In the toroidal model, although the entire tube is made buoyant, the `feet' of the flux tube are held at a fixed distance apart in the solar interior.  Figure \ref{footpoints} shows the $y$-separation of the maximum and minimum $B_z$ the base of the photosphere $(z=0)$ for the twist cases $\alpha=0.2,0.3,0.4$.  As described above, the lower the twist, the faster the rise to the photosphere.  Once the tubes reach the photosphere, the fields begin to spread.  The sunspot separation increases linearly until a peak distance is reached and the $y$-separation remains approximately constant.  This peak distance corresponds to the major diameter of the initial tube.  In this example, $2s_0=30$.  The constancy of the separation between the two main opposite polarities (sunspots) in an active region is often used as a criterion for the region's maturity \citep{liu06}. Tubes of a higher twist will spread laterally more slowly than tubes of a lower twist since a higher twist produces a stronger tension force.  The slopes for the linear separation phase are estimated to be 0.833 for $\alpha=0.2$, 0.815 for $\alpha=0.3$ and 0.8 for $\alpha=0.4$.

The maximum separation of the sunspots in these simulations is determined by the distance of the almost vertical flux tube legs at the base of the numerical box.  This has consequences for the structure of the magnetic field in the interior.  The classic picture of flux emergence, as described in the Introduction, considers flux tubes with long wavelengths in the convection zone, typically generated by $m=1, 2$ instabilities at the tachocline, where $m$ is the longitudinal wavenumber.  In the local Cartesian approximation near the surface,
this can be represented by the cylinder model.  The toroidal model, on the other hand, has vertical legs.  There are two possible mechanisms for forming toroidal tubes in the convection zone. Either the modification of the cylindrical tube takes the form of that described in $\S$ \ref{pd}, namely the (enhanced) buoyant region of the cylindrical tube is spatially limited and takes effect deeper in the convection zone (rather than near the
photosphere as in previous simulations) or the instability, at the base of the convection zone, involves a higher longitudinal mode,  e.g. $m>10$.  Both cases would produce toroidal shaped tubes with almost vertical legs.

\section{Conclusions}

In this paper we have carried out a parametric study of the emergence of buoyant \emph{toroidal} flux tubes through the solar interior and into the atmosphere, via 3D MHD simulations,.  By varying two of the parameters, namely $B_0$ and $\alpha$, we have been able to investigate the general behaviour of the emerging tubes.

Keeping $\alpha$ constant, the variation of the initial field strength produces a wealth of behaviour.  In the solar interior the buoyancy force is proportional to $B_0^2$.  Tubes with stronger $B_0$ rise faster and further than those with lower values.  In the solar interior the tubes exhibit a self-similar behaviour.  By rescaling the time to $\bar{t} = tB_0$, the axis height-time profiles for the different $B_0$ lie on top of each other \citep{murray06}.  This shows that the axis heights of the tubes are not only a function of time but also of $B_0$.

Once emerged, the evolution of the tubes can vary strongly depending on the choice of the initial field strength.  A value between $B_0=5$ and $B_0=7$ gives a threshold between two general classes of behaviour.  For the $B_0=7$ case, the axis rises fast enough to emerge into the atmosphere before plasma draining from the field above flows into a pressure deficit and blocks its ascent.  The plasma instead drains below the axis and produces reconnection upflows that further increase the height of the axis.  For the $B_0=5$ case, the axis does not rise fast enough to escape plasma draining on top of it and so is pinned down in the photosphere.

An advantage of the the toroidal model over the standard cylinder model in flux emergence is that the axis of the original tube is able to emerge into the atmosphere and so the field at the centre of the sunspots is vertical.  This is made possible by the ability of the the plasma to drain down the legs of the tube in the solar interior.  In the cylinder model such draining is not possible and plasma collects in dips along the tube axis.  It has been shown, however, that by changing the buoyancy profile to produce a toroidal shape, the tube axis can achieve greater heights in the solar atmosphere.

Keeping $B_0$ constant, the variation in $\alpha$ also produces interesting behaviour.  In the solar interior it is found that tubes with lower twists rise faster than tubes with higher twists.  Once at the photosphere, however, low twist tubes take longer to become subject to the magnetic buoyancy instability.  This allows higher twist tubes to catch up and emerge approximately at the same time.  

It is found that the amount of flux emerged into the atmosphere depends on the value of $\alpha$.  The higher the twist the more flux is transported upwards.  In the rise of a flux tube, the stronger the twist, the more preserved the tube remains.  The field strength of higher twisted tubes is stronger than lower twisted tubes when they emerge.  This manifests itself in the amount of flux that is transported into the atmosphere.

Another feature of the toroidal model, which improves upon the cylinder model, is that the sunspots drift to a fixed separation and then stop.  This separation is determined by the major diameter of the original tube.  The $y$-separation of the maximum and minimum values of $B_z$ increases linearly until the maximum separation is reached, the diameter of the tube.  The higher the twist of the the tube, the slower the separation rate.

This study has revealed many important aspects of the emergence of toroidal tubes and there are many directions for future work to follow, such as the inclusion of convection into the model to study how it affects the emergence and eventual break up of active regions and the interaction of multiple loops to model solar atmospheric events.

\begin{acknowledgements}
DM acknowledges financial assistance from STFC.  Part of the computational work for this paper was carried out on the joint STFC and SFC (SRIF) funded cluster at the University of St Andrews.  DM and AWH acknowledge financial support from the European Commission through the SOLAIRE Network (MTRN-CT-2006-035484).  DM and AWH would also like to thank the referee for helpful and constructive comments.
\end{acknowledgements}

\end{document}